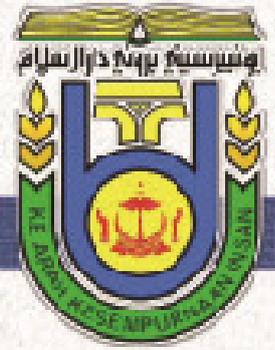

# FBEPS
# AGBEP
## PhD Colloquium

"Bridging the Research
Community Towards Excellence"

15-16 Rejab 1433 | 5th - 6th June 2012
The Core
Universiti Brunei Darussalam

ubd

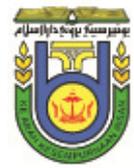

# Information Security Awareness
# Within Business Environment: An IT Review

**Heru Susanto[12] and Mohammad Nabil Almunawar[1]**

**[1]University of Brunei**
FBEPS, Information System Group
*susanto.net@gmail.com*
*nabil.almunawar@ubd.edu.bn*

**[2]The Indonesian Institute of Sciences**
Information Security &
IT Governance Research Group
*heru.susanto@lipi.go.id*

**Abstract -***The beauty of Information Technology (IT) is with its multifunction nature; it is a support system, a networking system, a storage system, as well as an information facilitator. Aided with their broad line of services, an IT system aims to support or even drive organisations towards desired paths. Trends of IT and information security awareness (ISA) in society today, particularly within the business environment is quite interesting phenomenon. The overviews of the role of IT in the modern world as well as the perception towards ISA are initially introduced. A series of scope are outlined, and also further examination on matter of IT and ISA in the business environment–emphasis on revolution of business with ISA, security threats such as identity thefts, hacking and web harassments, and the different mode of protections that are applied in different business environments. Unfortunately, the advancement of IT is not followed by the awarness of its security issues properly, especially in the context of the business settings and functions. This research and review is expected to influence the awarness of information security issues in business processes.*

## I. INTRODUCTION

IT shaped the success of organisations, giving a solid foundation that increases both their level of efficiency as well as productivity. Over the years, IT development has become a lucrative industry as demands for a better – and more innovative – IT system continue to rise. Nowadays, the era of modern technology where we hardly avoid IT in our daily life.People are becoming increasingly aware of the IT benefits,especially as these benefits are mostly in tune with the fast-paced nature of the society today. On the other hand, in supporting IT performance, information securityawareness over IT is extremely needed. Information security means protecting information and information systems from unauthorized access, use, disclosure, disruption, modification, perusal, inspection, recording or destruction.Governments, military, corporations, financial institutions, hospitals, and private businesses amass a great deal of confidential information about their employees, customers, products, research, and financial status. Most of this information is now collected, processed and stored on electronic computers and transmitted across networks to other computers.Should confidential information about a business' customers or finances or new product line, fall into the hands of a competitor? Such a breach of security could lead to negative consequences. Protecting confidential information is not only a business requirement, but also an ethical and legal requirements. Therefore, information security awareness (ISA) is the stakeholder knowledge and attitude within an organization toprotectedtheir physical andinformation assets. Several issues in ISA as follows:

- The nature of sensitive material and physical assets they may come in contact with, such as trade secrets, privacy concerns and government classified information.
- Employee and contractor responsibilities in handling sensitive information, including review of employee nondisclosure agreements.
- Requirements for proper handling of sensitive material in physical form, including marking, transmission, storage and destruction.
- Proper methods for protecting sensitive information on computer systems, including password policy and use of two-factor authentication.
- Other computer security concerns, including malware, phishing, social engineering, etc.
- Workplace security, including building access, wearing of security badges, reporting of incidents, forbidden articles, etc.
- Consequences of failure to properly protect information, including potential loss of employment, economic consequences to the firm, damage to individuals whose private records are divulged, and possible civil and criminal penalties

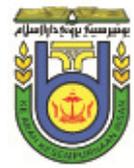

Being security aware means stakeholders understand that there is the potential for some people to deliberately or accidentally steal, damage, or misuse the data that is stored in a company's computer system. Therefore, it is crucialto protect the assets of the institution (information, physical, and personal) from any possible security breach. 'The focus of information security awareness should be to achieve a long term shift in the attitude of employees towards security, whilst promoting a cultural and behavioural change within an organisation. Security policies should be viewed as key enablers for the organisation, not as a series of rules restricting the efficient working of stakeholder's business.'

While most of IT systems are designedto have a considerable strength to assist corporateand organisations in operating their business processes or achieving their goals, unfortunately such systms are not immune from security threats (figure 1). Although the level of potential security threatvaries according to the nature of the threat as well as the system's ability to protect itself, the  existence of  risks attached to these systems is enough to rattle the level of trust that people have towards them.If risks are not reduced or threats cannot be neutralized, stakeholders –especially those who aware of security issues, hesitate to utilize IT systems offered for their activities, especially for electronic business activities.

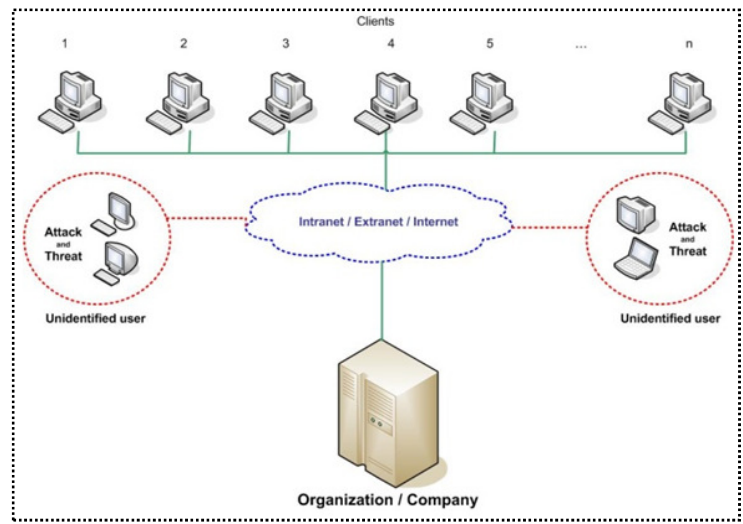

*Figure 1. Activities of unidentified user as potential attack and threat to organization*

IT systems currently   are becoming more prominent in business. In order to comprehend the level of prominenceof  IT systems over business environment, a series of review must be conducted. Furthermore, it is also important to understand how far the IT performancerevolutionize the business process within organisations and its effect on the efficiency as well as productivity.

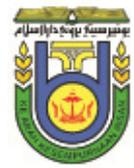

Our research is primarily aimed to  identify theorganization awareness level on information security issues. It is obvious that a different threat indicated a different risk level, thus, it is important to identify the various threats that are posed towards the IT system observed. This paper is organized as follows. In the section 2, we discuss several information technology issues and its contribution as agent of change in business. Issues and current trend in information security awareness is discussed in Section 3. Therefore, in same section an information security awareness breaches and its impact to successful business process within corporate and business environment. Section4, we discussed several standards for information security and proposed novel framework (ISF), based on ISF, we developed software to measuring level of information security awareness within business environment to protecting information and information systems from unauthorized access, use, disclosure, disruption, modification, perusal, inspection, recording or destruction. Finally, conclusion remarkand future work are provided in Section 5.

## II.    REVIEW: IT AS AGENT OF CHANGE IN BUSINESS PROCESSES

The presence of IT in business has gradually become more important as more business corporateintegrated their business processes with IT. An IT system can help business in various aspects including a platform for businesses such as storingand retrieving information, work synchronization, forming business strategy and communication.  The importance of IT in business is highly recognized (Solis, 2012) it is important for a corporation to seek an initiative to understand the mechanisms behind the system. The information technology revolution has altered the entire business world (Susanto et al, 2012b), it has dramatically revolutionized the mechanism of business process within and between corporate in supporting their activity.

There are two interesting hypotheses– both of whichrevolvearound how information revolution intensified (Porter & Millar, 1985). The first hypothesis stated an evolution of information technology allows the complete alteration of company operation, reshaped their product and outputs completely, indicatedby laying out of two significances strategic; the first through '*value chain*' – this concept measures the role of IT in shaping competition – and the second is through '*value activities*' – which revolved completely around the willingness of consumers to engage with, by provided a clear picture on how the IT completely permeate the value chain consecutively, whereby the improvement in the information-processing component in the system, the rate of data capture and manipulation became faster. Hence this subsequently enlarged the data channel, and in turn reduced the cost as well as increases the efficiency of information processing which was previously accomplisedsolely by human labour. Furthermore, the evolution of the IT deflated the limitations of what business companies think they are capable of, and instead allowed them to reach new potentials – it is the reason why competition in the business environment intensified which is elaborated.

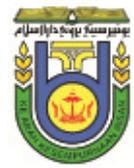

The second hypothesis refers on how IT changed rules and the nature of the competition in the business environment. Here, competition as a necessity to improve and provide a clearer path towards corporate's direction. The revolution of information technology created a chain reaction, whereby the vast improvements in the IT system drove corporate to use them as a lever in competitions, and would subsequently alters the structure of the industry as other firms tried to imitate the success of the leading, and itwould finally lead to the development of new business. Therefore, Nexis – an electronic data base company –their fast and efficient service enable user to substitute several noted library research as well as consulting firms. Nexis allowed subscribers to reduce their time when didliterature searches, it indicatednot only reduce unnecessary time consumption but also helps reduce cost as Nexis only requested payment for the specific information required without forcing subscribers to fully subscribe to the journals – unlike other library research corporates. The second point is the creation of competitive edge; whereby as Nexis successfully developed a system which is both fast and efficient for texts and periodicals search, and furthermore it was also relatively innovative during that time period which differentiate from their peers, such gave them a significant competitive edge. As Nexis was able to meet the demands of students, researchers or even other consulting firms in a faster rate, thus this significantly affected their loyalty and then established company market share in related market segmentation.

The final point on how information technology spawned new potentials business, which mainly implemented through technology feasibility, it to create demands for new products (Porter & Millar, 1985). 'Zapmail' by Federal Express as an example, whereby IT allowed combination betweensensing, imaging and telecommunications to spurred a new technology facsimile service. On other hand, Western Union provided 'Easy-Link' service to the public as their main business in provided high-speed data communication network. The surge of IT as a movement towards early globalization, as it increased demand for society as a whole to have an efficient and fast data service, increasing their needs for a better communication system not just limited regionally but globally.

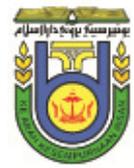

However, when we look an application of IT to the 21st century, Porter and Millar (1985) indeed gave a correct prediction as IT continues to be the force driverin shaping the business environment today. Furtehmore, Furnell and Karweni (1999) argue that the Internet is one of the methods used by businesses to communicate with clients as well as with the employee. This business operation method is known as electronic commerce. The speed, low cost, dependable and accessible features of the Internet are considered as the main reasons as to why most firms embrace electronic business and electronic commerce.

Based on a survey in Ameritrade Holding Corporation (Furnell & Karweni, 1999), it is believed that businesses on the Internet increased to 14.4 million by 2002 from approximately 3 million of businesses on that year– it indicatedthat business and corporate were prosper when they used IT in supporting their business processes. Nowadays, over 103 million business running and operating thorough the Internet (whois.sc, May 25, 2012), and also there are over 3.146 billion of active email accounts worldwide. Furthermore, it is found that due to the increase of profit margin business conducted through the Internet, there is an increase demand to strengthen security and protection of data and systems. Trust is one of the greatest elements that customers and stakeholderlook upon. Surprisingly, in the survey it is found that only 33 per cent of the respondents are aware of the related issues of trusted third parties while the majority – 80 per cent of them – are only concerned about the Data Encryption Standard (DES- is a previously predominant algorithm for the encryption of electronic data). Therefore, the issue of information security awareness is still important to address. Secondly, a survey was distributed to business sectorbased in the United Kingdom (UK). Unfortunately, as the result, their security awareness is quite low as it is found that there was poor of concern on the security requirements whereby the viruses and system unavailability were considered to be minor concern for these firms (Furnell and Karweni, 1999).

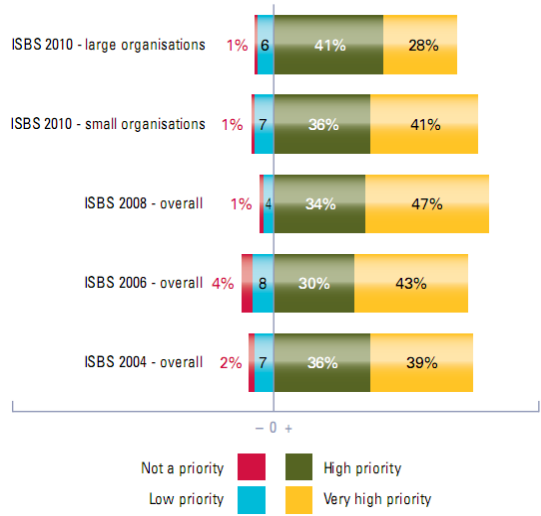

**Figure 2.** *How high a priority is information security to top management or director groups? (ISBS, 2010)*

Fortunately, the information security breaches survey (2010) indicated the need to set the tone at the top is a common from information security professionals. 77% of top management –small organization and 69% of top management –large organizationgive a high priority to security (figure 2). Government, financial services and technology organisations assign the highest the priority to security.Retailers are another outlier, with over a quarter indicating that their security is a low priority.

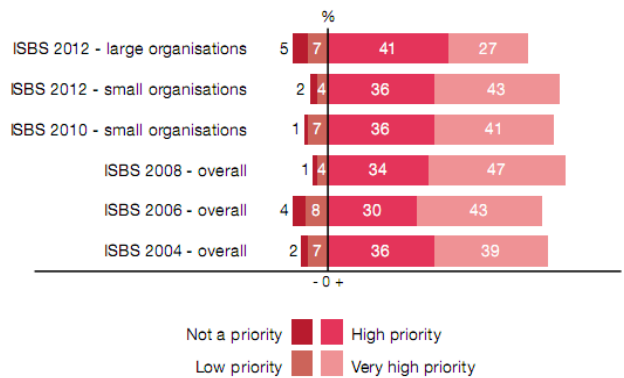

**Figure 3.** *How high a priority is information security to top management or director groups? (ISBS, 2012)*

Senior management or board of management support is vital if staff are to manage security effectively. So, it is encouraging that three quarters of respondents believe security is a high or very high priority to their senior management. In contrast, one in eight IT and information security personnel feels security is a low priority. However, the highest priority was reported by retail and distribution firms, twice as high as for property and construction companies. 79% of top management –small organization and 68% of top management –large organization give a high priority to security (figure 3), some respondents in large organisations were damning about the lack of priority they see and the impact has (ISBS, 2012).

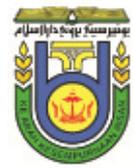

## III.   IMPACT OF ISA BREACHES

However, the world is changing, the technology is developing as well. Innovation in technology caused the number of hackers to grow too. The changing business environment is creating new vulnerabilities. Criminals are adapting their techniques exploiting the vulnerabilities. as a result cybercrime is becoming more common. The number and cost of security breaches appears to be rising fast (ISBS, 2010). In 'Human Resource Management International Digest' (2005), stated communication company offering free security-awareness training to motivated customers. It focuses on assisted stakeholder to understand the various security threats and also provide measurement to avoid unnecessary losses due to security failure. They concerned in fact that only one company in ten has staff with formal information-security qualifications, as the root cause of security failure in an organisation.

Therefore, shown that 70% of IT budget are spent after experiencing security breach. To combat, Energis is partnering with ThombsonNET; a provider of corporate education and training, to provide the online security training which also offers consultancy and coaching services from Energis' experts (Pollitt, 2005). They are providing new ways to prevent customers from security threats, through concept of "five principles of security"; Planning, Proactive, Protection, Prevention and Pitfalls*, to identify signs and flags of intruders in a network, establish guidelines for safeguarding user names and passwords, match protection against physical access to information-technology facilities with the level of threat and the security systems which are most effective to deal with.

The interaction of Actor Network Theory (ANT) and the Due Process Model with information security management in a specific organisational context, understand and manage security awareness activities,indicated that ISA activities need the participation of both human actors and non-human actors. ANT has been created by the Science and Technology Studies (STS) by Latour (1996), the main purpose is to address the role of technology in a social setting and analyse the processes by which technology affects by the social element of a context. ANT outlines how actors (e.g. top managers, administrators, security officers, users, programmers, Perpetrators) form alliances and enrol other actors, by using non-human actors (e.g. the information security policy or the information security plan/programme, relevant standards), to strengthen such associations and their interests.

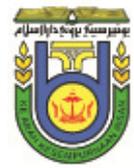

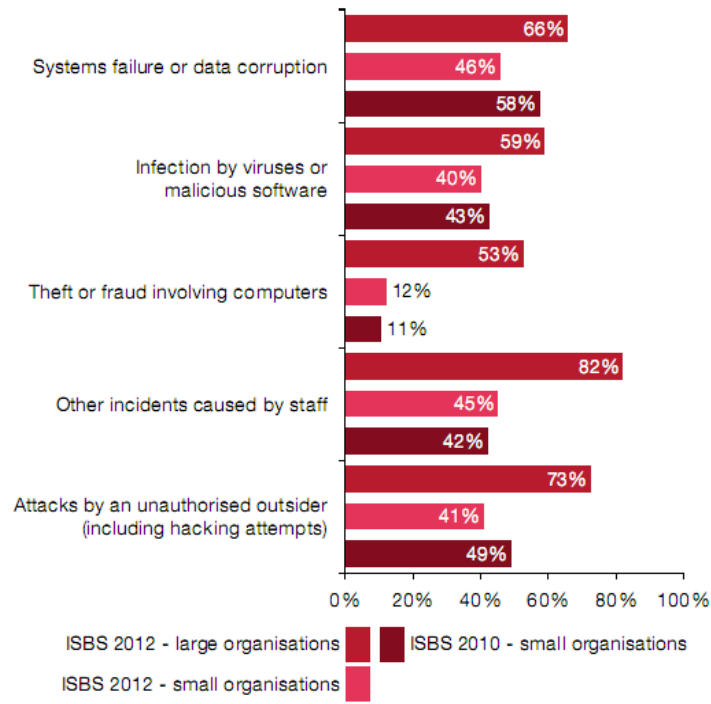

***Figure 4.*** *What type of breaches did respondents suffer? (ISBS, 2012)*

Furthermore, ISBS (2012) stated that 'incidents caused by staff' contirbuted 82% for large organization. It indicated internal disturbance caused by staff is quite increase from 42% by 2010.No industry sector appears immune from these incidents. Telecommunications, utilities and technology companies appear to have the most reliable systems. The public sector and travel, leisure and entertainment companies are most likely to have systems problems (figure 4).

.

Unfortunately, It is found that the average businesses have estimated at least one security incident a month while large companies would expect once a week. The average cost of an organization's most serious security incident is £10,000 - £20,000 by 2008, increased to £27,000 - £55,000 by 2010. In other hand for large companies the overall cost of an organisation's worst incident is £120,000 - £120,000 in 2008 and it is grow up to £280,000 - £690,000 among 2010. Therefore in Feb-Mar 2012, for small companies£15,000 - £30,000 and £110,000 - £250,000 for large companies (tabel 1).

| | ISBS 2012 - small organisations | ISBS 2012 - large organisations |
|---|---|---|
| Business disruption | £7,000 - £14,000 over 1-2 days | £60,000 - £120,000 over 1-2 days |
| Time spent responding to incident | £600 - £1,500 2-5 man-days | £6,000 - £13,000 15-30 man-days |
| Direct cash spent responding to incident | £1,000 - £3,000 | £25,000 - £40,000 |
| Direct financial loss (e.g. loss of assets, fines, etc.) | £2,500 - £4,000 | £13,000 - £22,000 |
| Indirect financial loss (e.g. theft of intellectual property) | £4,000 - £7,000 | £5,000 - £10,000 |
| Damage to reputation | £100 - £1,000 | £5,000 - £40,000 |
| Total cost of worst incident on average | £15,000 - £30,000 | £110,000 - £250,000 |
| 2010 comparative | £27,500 - £55,000 | £280,000 - £690,000 |
| 2008 comparative | £10,000 - £20,000 | £90,000 - £170,000 |

*Tabel 1. What was the overall cost of an organisation's worst incident in the last year? (ISBS, 2012)*

Cloud computing is coming, which offering organizations in three types of services; software-as-a-service, infrastructure-as-a-service and platform-as-a-service. Cloud computing is improving security. But many want better enforcement of provider security policies, among other priorities. Has the cloud improves security? More than half (54%) say it has, 23% believe that security has "weakened" and 18% see no change (*GSISS, 2012*) (figure 5).

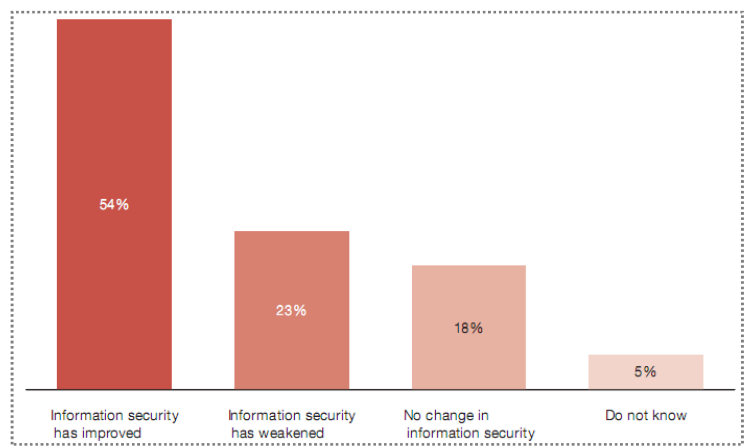

*Figure 5. The Impact of cloud computing on information security (GSISS, 2012)*

Chris Potter (ISBS, 2012), said "*The UK is under relentless cyber-attack and hacking is a rising risk to businesses. The number of security breaches large organizations are experiencing has rocketed and as a result, the cost to UK plc of security breaches is running into billions every year. Since most businesses now share data with their business partners across the supply chain, these numbers are startling and make uncomfortable reading for business leaders.Large organizations are more visible to attackers, which increases the likelihood of an attack on their IT systems. They also have*

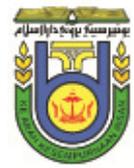

*more staff and more staff-related breaches which may explain why small businesses report fewer breaches than larger ones. However, it is also true that small businesses tend to have less mature controls, and so may not detect the more sophisticated attacks."*

Despite the prolonged economic slowdown, most organizations have spent more on security this year than in the previous one. On average, organizations spend 8 percent of their IT budget on information security, and those that suffered a very serious breach were found to spend on average 6.5 percent of their IT budget on security. However, that it is hard to measure the business benefits from spending money on security defences. Only 20 percent of large organizations evaluate return on investment on their security expenditure Chris Potter (ISBS, 2012).

## IV.   ISA AND ISF WITHIN BUSINESS ENVIRONMENT

Information security awareness that concernson IT applications especially email breaches has prompted businesses to review how employees handle email, as well as the role of technologies, policies and user education in determining how an email systemshould safely deployed. External threats such as hackers and viruses need to be addressed through the use of technology; firewalls and data encryption must be effectively deployed. It is important to note that the usage of locally technical solutions such as scanning and monitoring software is extremely required (Susanto, Almunawar and Tuan, 2012a). Jago (2001) found the correspondences that are sent by employees via email are the responsibility of companies. Finally, email security is an important issue in the current business climate and basically three important elements might be used to improve  email security technologies, explicit policies and user education.

Business Information Review (2005) stated in how an email system developed from its early days as an efficient and cost-effective way of communicating short messages across computer networks. Consequently, it has become a critical business application for various purposes other than merely sending-receiving  information. The report also stated that the email applicationis as an integral part of daily business. Other related applications such as a scheduling tool and an address book  have been integrated to the email application   and they are utilized by employees to improve their productivity (Kelleher & Hall, 2005). In addition, the private credit card transactions were found to be insecure from misuse and personal information such as addresses, and telephone numbers. The administration of electronic commerce security is defenceless, most threats are unresolved and illegal activities such as eavesdropping, password sniffing, data modification, spoofing and repudiation are becoming more frequent. The growth of electronic commerce in an organization, as well as the possibility of fraud and dishonesty happens broadly in the marketplace, where the hackers could access information of credit card and confidential information in fraudulent activities more easily. The authentication and policies over information security would prove who the parties involved in the electronic transaction and communications are hence would ensure the integrity of data on the system. However, It is imperative for organizations to use an information security management system (ISMS) to effectively manage their information assets. ISMS is basically consist of sets of policies put place by an organization to define, construct, develop and maintain security of their computer based on hardware and software resources. These policies dictate the way in which computer resources can be used.

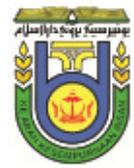

Since information security has a very important role in supporting the activities of the organization, we exteremely need a standard or benchmark which regulates governance over information security, these policies and standard as function as fundamental guidelines forcorporate secure electronic commerce on the global scale (Susanto et al, 2011b, 2012b). There are several standards for IT Governance which leads to information security awareness such as PRINCE2, OPM3, CMMI, P-CMM, PMMM, ISO27001, BS7799, PCIDSS, COSO, SOA, ITIL and COBIT. Unfortunately, some of these standards are not well adopted by the organizations, with a variety of reasons. The big five of ISMS standards, ISO27001, BS 7799, PCIDSS, ITIL and COBIT, widely used standards for information security is discussed and compared (figure 4) (Susanto, Almunawar, and Tuan 2012a)to determine their respective strengths, focus, main components and their adoption based on ISMS.

In conclusion (Susanto, Almunawar, and Tuan 2012a) each standard playing its own role and position in implementing ISMS, several standards such as ISO 27001 and BS 7799 focusing on information security management system as main domain and their focus on, while PCIDSS focus on information security relating to business transactions and smart card, then ITIL and COBIT focuses on information security and its relation with the project management and IT Governance (figure 6). Refers to the usability of standards in global, indicated that ISO (27001) leading than four other standards especially on ISMS, therefore it indicated that the standard is more easily implemented and well recognized by stakeholders (top management, staff, suppliers, customers/clients, regulators), the standard introduces a cyclic model known as the "Plan-Do-Check-Act" (PDCA) model, aims to establish, implement, monitor and improve the effectiveness of an organization's ISMS (Susanto, Almunawar & Tuan 2011b), thus compliance with information security standard, ISO 27001, is highly recommended with a variety of reason mentioned.

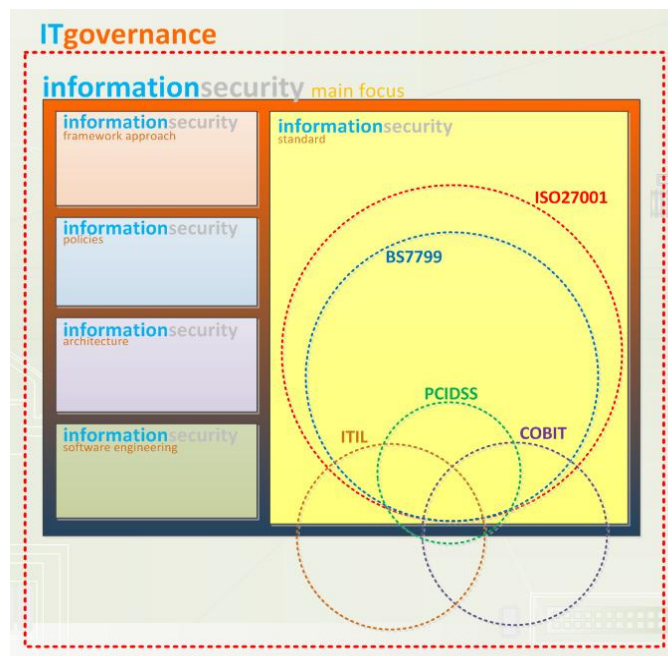

***Figure 6**. Position of information security standard*

Therefore, ISO 27001 is designed to assure the confidentiality, integrity and availability of information assets. It is exclusive for information security, and only addresses that issue (Alfantookh, 2009). The key areas identified by ISO27001 for the implementation of an information security management system are;

1. ***Information Security Policy:*** how an institution expresses its intent with emphasized to information security, means by which an institution's governing body expresses its intent to secure information, gives direction to management and staff and informs the other stakeholders of the primacy of efforts.

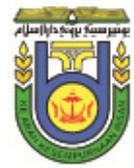

2. ***Communications and Operations Management:*** defined policy on security in the organization, in reducing security risk and ensuring correct computing, including operational procedures, controls, and well-defined responsibilities.

3. ***Access Control:*** is a system which enables an authority to control access to areas and resources in a given physical facility or computer-based information system.

4. ***Information System Acquisition, Development and Maintenance:*** an integrated process that defines boundaries and technical information systems, beginning with the acquisition, and development and the last is the maintenance of information systems.

5. ***Organization of Information Security:*** is a structure owned by an organization in implementing information security, consists of; management commitment to information security, information security co-ordination, authorization process for information processing facilities. Two major directions: internal organization, and external parties.

6. ***Asset Management:*** is based on the idea that it is important to identify, track, classify, and assign ownership for the most important assets to ensure they are adequately protected.

7. ***Information Security Incident Management:*** is a program that prepares for incidents. From a management perspective, it involves identification of resources needed for incident handling. Good incident management will also help with the prevention of future incidents.

8. ***Business Continuity Management:*** to ensure continuity of operations under abnormal conditions. Plans promote the readiness of institutions for rapid recovery in the face of adverse events or conditions, minimize the impact of such circumstances, and provide means to facilitate functioning during and after emergencies.

9. ***Human Resources Security:*** to ensure that all employees (including contractors and user of sensitive data) are qualified for and understand their roles and responsibilities of their job duties and that access is removed once employment is terminated.

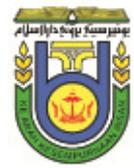

10. ***Physical and Environmental Security:*** to measures taken to protect systems, buildings, and related supporting infrastructure against threats associated with their physical environment, buildings and rooms that house information and information technology systems must be afforded appropriate protection to avoid damage or unauthorized access to information and systems.

11. ***Compliance:*** these issues necessarily are divided into two areas; the first area involves compliance with the myriad laws, regulations or even contractual requirements which are part of the fabric ofevery institution. The second area is compliance with information security policies, standards and processes.

## PROPOSED FRAMEWORK -ISF

In this subsection we proposed novelty approach and new paradigm on recognizing an organization awareness level - information security issue – ISA, by introducing a new framework which leads to the development of mathematical models as novel approach for information security compliance measurement of organization readiness level, called by I-SolFramework (ISF) (figure 7) abbreviation from Integrated Solution for Information Security Framework (Susanto, Almunawar and Tuan, 2012c, 2012d). Therefore, the framework designed to handle different types of organizations and standards (Susanto et al, 2011c). It consists of six major domains namely: Organization, Stakeholders, Tools & Technology, Policy, Culture and Knowledge. Elucidation of the term and concept is follows; *Organization:* A social unit of people, systematically structured and managed to meet a need or to pursue collective goals on a continuing basis, the organizations associated with or related to, the industry or the service concerned. *Stakeholder:* A person, group, or organization that has direct or indirect stake in an organization because it can affect or be affected by the organization's actions, objectives, and policies. *Tools & Technology:* the technology upon which the industry or the service concerned is based. The purposeful application of information in the design, production, and utilization of goods and services, and in the organization of human activities, divided into two categories (1) Tangible: blueprints, models, operating manuals, prototypes. (2) Intangible: consultancy, problem-solving, and training methods. *Policy:* typically described as a principle or rule to guide decisions and achieve rational outcome(s), the policy of the country concerning the future development of the industry or the service concerned. *Culture:* determines what is acceptable or unacceptable, important or unimportant, right or wrong, workable or unworkable. *Organization Culture:* The values and behaviors that contribute to the unique social and psychologicalenvironment of an organization, its culture is the sumtotal of an organization's past and currentassumptions (Alfantookh, 2009). *Knowledge:* in an organizational context, knowledge is the sum of what is known and resides in the intelligence and the competence of people. In recent years, knowledge has come to be recognized as a factor of production (BDO, 2012).

In addition, framework introduces refinement in order to determine the degree of clarity of each essential control over the parameters ISO 27001. Without a refined however, the controls tend to be somewhat disorganized and disjointed (Susanto et al, 2010b, 2011b). Finally, the ISO27001 information security standard mapped by related six-domains to the assessment issue and controls with posibly external attack and distruptsion phenomenon.

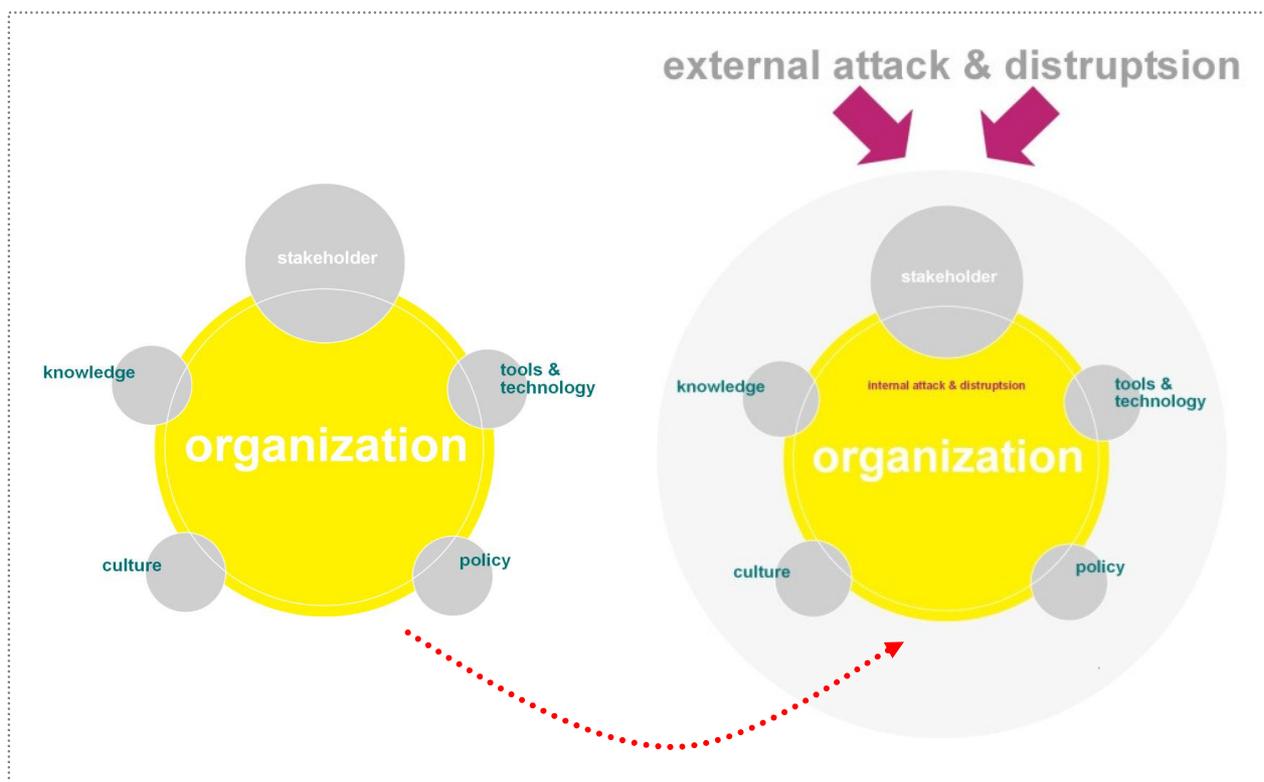

***Figure 7.*** *Proposed framework*

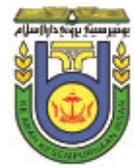

ISF's mathematical model describes the detail how computation is performed, and how to calculate the six domains, a results as function as an indicator of organization awareness level - information security issue.Mathvariables are defined, where $k$ as *control*, $j$ as *clause*, $i$ as *domain* and $h$ as a *top domain*, all those are ingredients of ISO 27001 mapped by ISF, details of these models as follows (Susanto & Almunawar, 2012d);

$$(1) \rightarrow x_h = \sum_{i=1}^{n} \frac{\left[\sum_{j=1}^{n} \frac{\left[\sum_{k=1}^{n} \frac{[control]_k}{n}\right]_j}{n}\right]_i}{n}$$

Therefore, for 6 domains of ISF (Organization, Stakeholder, Tool and Technology, Knowledge, Culture, and Policy), Eq. (1) can be expressed as:

$$x_h = \sum_{1}^{6} \frac{\left[\sum_{i=1}^{n} \frac{\left[\sum_{j=1}^{n} \frac{\left[\sum_{k=1}^{n} \frac{[control]_k}{n}\right]_j}{n}\right]_i}{n}\right]}{6}$$

## ISM for Measuring ISA Level within Organization

In this subsection we developed amodeling software, namelly integrated solution modelling software (ISM),as implementation of mentioned framework (ISF), as function as measuring information security awareness (ISA) level within organization and business environment, such as protecting information and information systems from unauthorized access, use, disclosure, disruption, modification, perusal, inspection, recording or destruction. ISM is consistsoftwomajorsubsystemsofe-assessment and e-monitoring. E-assessment is to measure ISO 27001 parameters based on the proposed framework with 21 essential controls and e-monitoring is to monitor suspected activities that may lead to security breaches. The software is equipped with a login system, as user track record, that can be used to determine users' patterns of assessment. Therefore, stakeholder is prompted to entering anachievements value based onISO27001 parameters,called by *assessment issues*. Level of assessment set out in range of 5 scales;

- ❖ *0 = not implementing*
- ❖ *1= below average*
- ❖ *2=average*
- ❖ *3=above average*
- ❖ *4=excellent*

As a measurement example, we described it in details to several steps of parameters assessment as follows (figure 8):

1. *Domain*: "***Organization***"
2. *Controls*: "***Organization of information security: Allocation of Information Security Responsibilities***"
3. *Assessment Issue*: "***Are assets and security process Cleary Identified?***"
4. Then stakeholders should analogize ongoing situation, implementation, and scenario in organization, and benchmark it to the security standard level of assessment as reference standard.

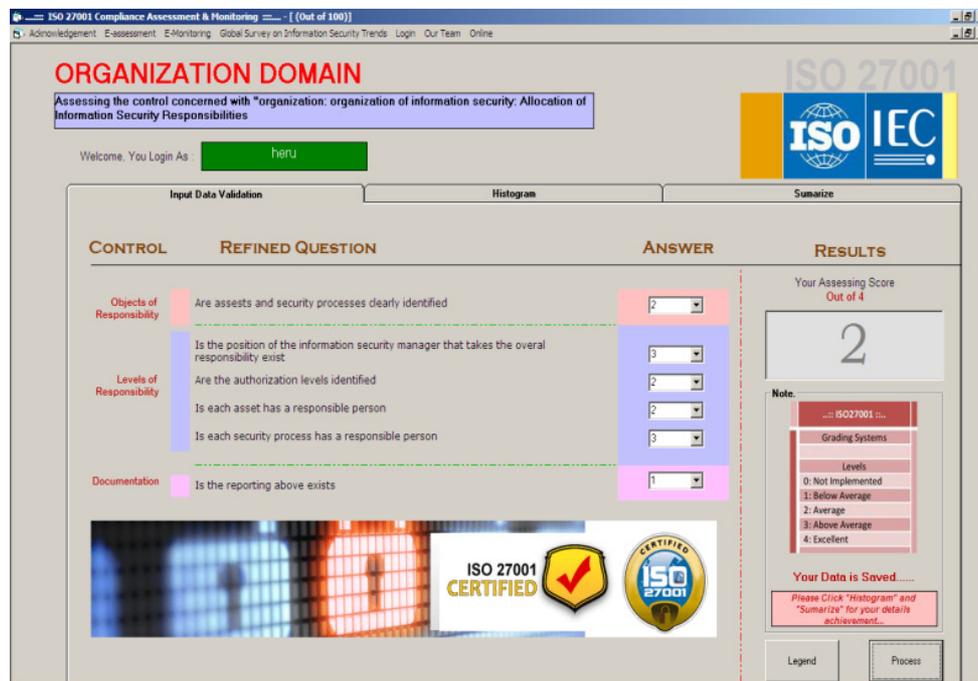

***Figure 8***. *Assessment form*

Histogram features (Figure 9 & 10) showed us details of the organization's achievement and priority. Both statuses reviews of strongest and weakness point on an organization current achievement. As indicated is the system, *"Achievement"* declared the performance of an organization as final result of themeasurement validated by proposed framework-ISF. Then,*"Priority"* indicated the gap between ideal values with achievement value. If achievement is high, then domain has a low priority for further work, and conversely, if achievement is low, then the priority will be high (figure 9). However, by estimated each parameter's performance, as an assessment and forecasting approach, stakeholder has a comprehensive overview on their current security awarenes level.

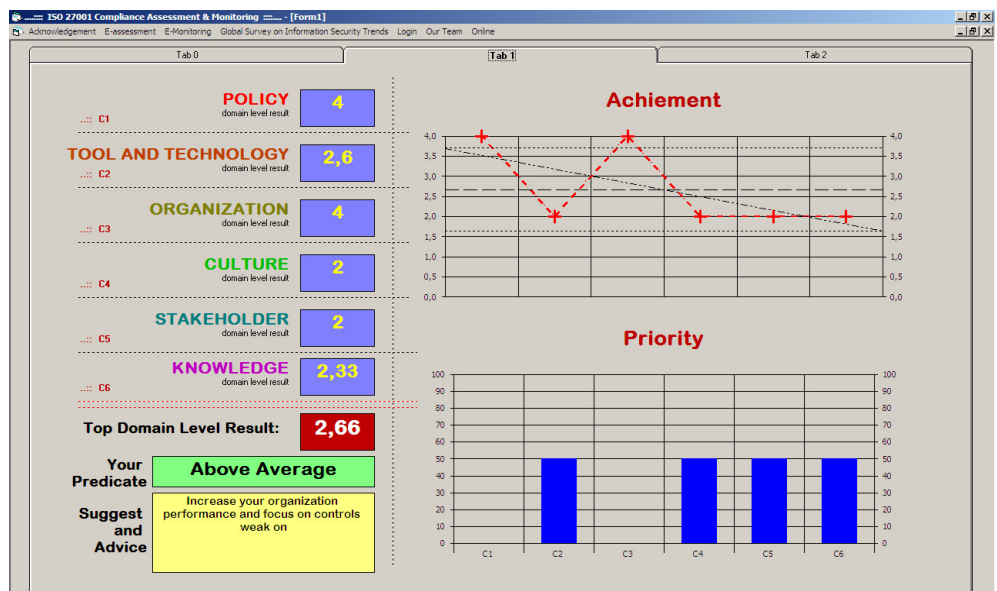

*Figure 9. Six domain final result view on histogram style*

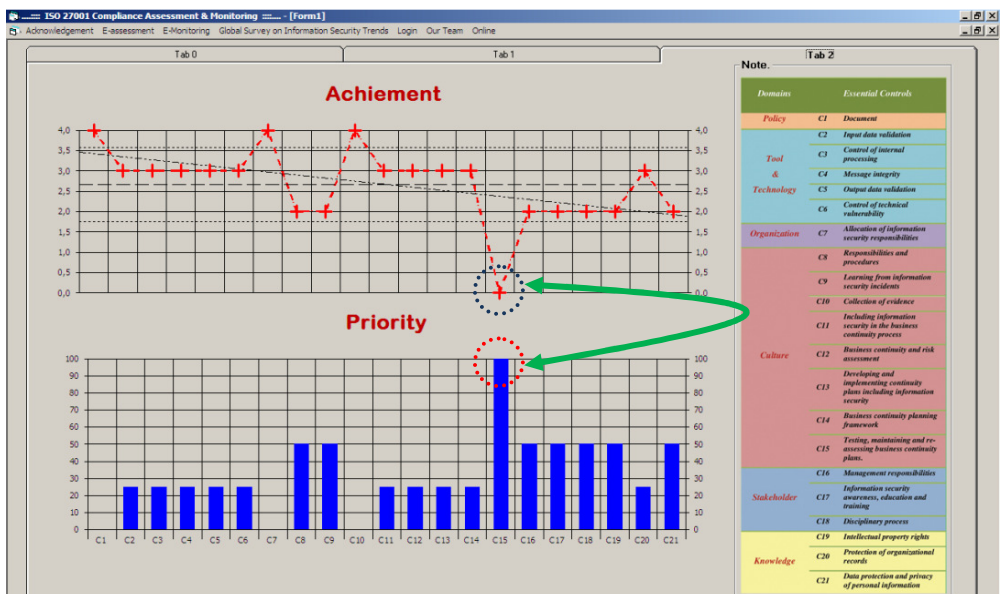

*Figure 10. 21 –essential controls final result view on histogram style*

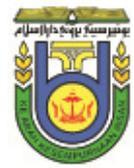

Therefore, we provided an illustrative measurement asindicatedin the following figures. Figure 9, illustrates the state of six domain controls related to top domain result. Figure 10, represents the condition of 21 essential controls of standards. The overall score of all domains is shown at "2.66" points. The domain of the "policy" and "organization" scored highest at "4", and the domain of the "culture" and "knowledge" scored lowest at "2". Achievement and priority figures are given to illustrate the strongest and weakness in the application of each control, c15 is lowest scored within 21 essential control of ISO 27001, it indicated as "*testing, maintaining and re-assessing business continuity plans*" (figure 10).

## V.   CONCLUSION REMARK

Information technology played a major role in changing how organizations, people and businesses interact with each other. The changes have impact both advantages and disadvantages. New segmentation is created, growth of new markets, international investment and trading are also increasing as influences of advancement in IT implementation. As take a look at globally today, most organizations being dependant on IT, therefore the risk of IT could not be avoided, it's essential to lessen the exposures by developing information security plan. Security experts need to be aware that the risk could possibly come even from their employees and customers. Therefore, security plans needs to carefully look at the people, policy and the technology in order for achieve its goal.

Most businesses need to take full advantage of technology. Speed, accuracy, competition, productivity and convenience are elements that businesses should look upon to expand their products and services. For example, online security - attracting tech-savvy clients could increase the flow of business, and would encourage user through online commerce.

## VI.   FUTURE WORK

Actually business should not only rely solely on the benefits of using technology; they should look at the quality and the content of their work. It is important for them to work together with the IT system - combining both their brainpower as well as the power of IT.

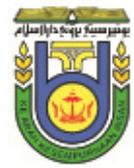

Therefore, with IT advancement and increasing dependency on IT, it is important that organisation's security system also go the same direction. More investment in the security system is essential for today's organisation in order to avoid security breach and protect information from unauthorised access. Furthermore, employee training for security awareness needs to be made compulsory and IT security expert should also be available in all business corporations. In the future it is important to place an emphasis on basic IT security skill – whereby it needs to be part of people's general knowledge and every employee who uses computer should know how to secure the information they are accessing and can identify intruder in the network. Security awareness activities should be implemented in all organisations. New ways to prevent security breach needs to be created and updated continuously in order to ensure organisation's assets are secured without possibility of leakage.

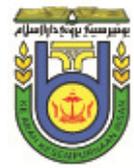

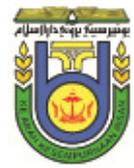

## AUTHORS

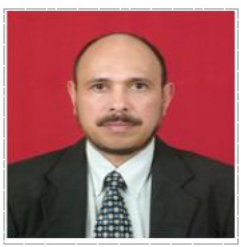

Mohammad Nabil Almunawar is a senior lecturer at Faculty of Business, Economics and Policy Studies, University of Brunei Darussalam. He received master Degree (MSc Computer Science) from the Department of Computer Science, University of Western Ontario, Canada in 1991 and PhD from the University of New South Wales (School of Computer Science and Engineering, UNSW) in 1997. Dr Nabil has published many papers in refereed journals as well as international conferences. He has many years teaching experiences in the area computer and information systems.

**Papers & Citations:** *http://scholar.google.com/citations?user=AJy07UIAAAAJ&hl=en*

Heru Susanto is a researcher at The Indonesian Institute of Sciences, Information Security & IT Governance Research Group, also was working at Prince Muqrin Chair for Information Security Technologies, King Saud University. He received BSc in Computer Science from Bogor Agriculture University, in 1999 and MSc in Computer Science from King Saud University, and nowadays as a PhD Candidate in Information Security System from the University of Brunei.

**Papers & Citations:** *http://scholar.google.com/citations?hl=en&user=rGgCYCoAAAAJ*

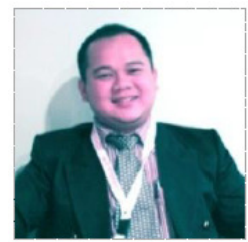

# FACULTY OF
## BUSINESS, ECONOMICS
### A N D   P O L I C Y   S T U D I E S

## UNIVERSITI BRUNEI DARUSSALAM

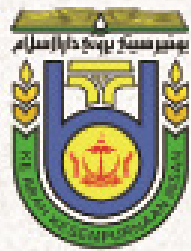

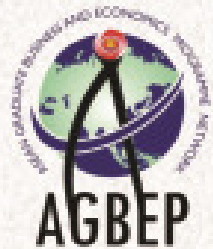

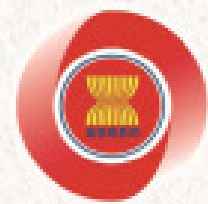